\documentclass[11pt]{article}
\usepackage{amsmath,amssymb,graphicx,hyperref,physics,listings,flafter}
\begin{document}

\begin{center}
Thermal convection in a Hele-Shaw cell with the dependence of thermal diffusivity on temperature
\end{center}

\begin{tiny}
a) R. C. Nuckchady,b) V. A. Demin \\

a) University of Oxford, UK email: rohan.nuckchady@gmail.com\\

b) Perm State University, email: demin@ psu.ru\\
\end{tiny}

\textbf{Abstract}\\
\\
The investigation of thermal convection of a fluid with the dependence of thermal diffusivity on temperature in a vertical Hele-Shaw cell heated from below has been fulfilled theoretically. The expression for equilibrium temperature distribution in a cavity has been derived analytically. It has been found that the dependence of temperature on the vertical coordinate looks like a square root law. The linear stability of mechanical equilibrium state against small normal perturbations has been investigated by means of Galerkin method. It has been shown that the most dangerous perturbation in a cavity under consideration is described by the mode which corresponds to the two-vortex steady flow. The numerical simulation of over-critical steady and oscillatory flows has been carried out in the approximation of plane trajectories. This simplification of theoretical model is consistent with all experimental data on thermal convection in similar cavities. It has been shown that the inclusion of the dependence of thermal diffusivity on temperature into the mathematical model leads to the "updown" symmetry breakdown for the small values of over-criticality.


\section{Introduction \\ 1.1. Thermal convection in the cavities of different shape}

In experiments, the evolution of the spatial pattern of convective flows is mainly determined by the shape of the cavity and the type of heating induced. In the case of a model infinite horizontal layer, steady flow has the cellular form and the vortices size is comparable with the thickness of the layer. But in a closed tank there will be a discretisation of allowed wavelengths and flow structure is strongly limited by the shape of a cavity. The type of boundaries also play an important role: the stability of a horizontal layer with free boundaries is fundamentally different from one with rigid boundaries [1]. Also, specific flows arise in cavities due to the particular thermal conditions or the presence of any spatial symmetry.

\subsection{Dependence of the kinetic coefficients on thermodynamics parameters}

In fluid dynamics the coefficient of thermal conductivity and the coefficient of viscosity are usually treated in most calculations as constants. While both theoretically and experimentally, these coefficients can be shown to depend on thermodynamic parameters, the dependence of those coefficients on temperature or pressure is usually neglected as these properties are more often insignificant. For example, the dependence of viscosity on pressure may be completely ignored when dealing with an incompressible fluid. However in the presence of a significant temperature gradient, the dependence of those coefficients on temperature may become substantial.

In $[2,3]$ the temperature dependence of viscosity was taken into account in order to explain the up-down symmetry violation of steady and oscillatory convective flows in a vertical Hele-Shaw cell heated from the bottom. Thus in the presence of strong environmental conditions those dependencies can not be trivially ignored.

\subsection{Dependence of thermal diffusivity} on temperature

In the analysis of most of convective problems, the coefficient of thermal diffusivity $\chi$ is assumed to be constant. In reality, it can depend both on temperature and pressure from general point of view. However, experiments have shown that the dependence of thermal diffusivity on pressure is much less significant than its dependence on temperature. While extremely small value the dependence on pressure can be neglected, the extent of the effect of temperature is less clear. In particular, the dependence of thermal diffusivity on temperature could help explain the symmetry breakdown in convective flows in Hele-Shaw cell.

Free thermal convection in a vertical Hele-Shaw cell has been studied thoroughly for a long time [4]. It was shown experimentally and mathematically that the convection threshold is determined by a steady flow in a cavity heated from the below. The number of vortices is governed by the degree of the cell elongation along horizontal axis. However the effect of the temperature dependence of thermal diffusivity on the boundary of stability and non-linear convective regimes has not been examined yet. In particular the stability of these states with $\chi(T)$ has not been studied in detail and the effect could be observable and measureable in an experiment

\section{Statement of the problem}

\subsection{Hele-Shaw cell}

Let us consider rectangular cavity with specific aspect ratio when the height and length are much greater than the thickness (see Fig. 1). It means mathematically that $h, l>>d$. This cavity is called by the HeleShaw cell [4]

The cavity is filled by a fluid with a temperature dependent thermal conductivity. The Hele-Shaw cell is heated from the bottom, $\Theta$ is the temperature difference between upper and lower heat exchangers. The approximation of plane trajectories is valid for a wide range of the governing parameters. Even chaotic regimes for large values of over-criticality visually have no transverse component of velocity $\left(v_{z}=0\right)$ and the trajectories of liquid particles lie in the $(x, y)$ plane.
\begin{center}
\includegraphics[scale=0.5]{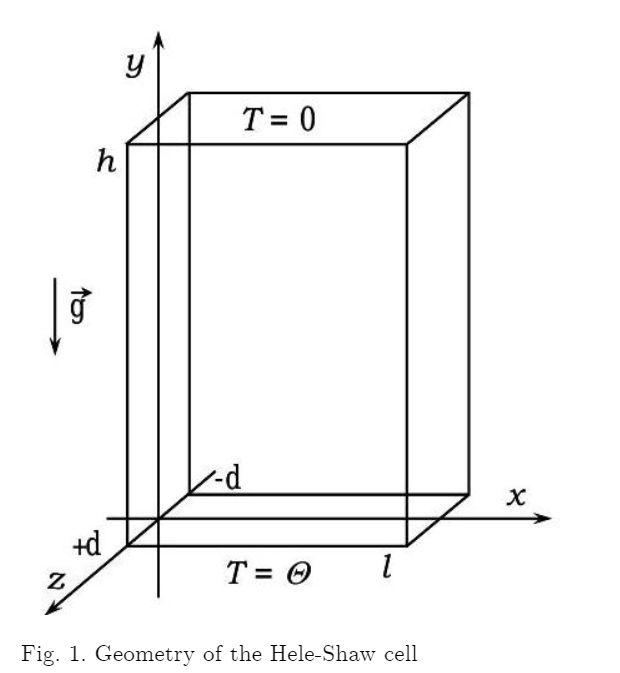}
\end{center}
\subsection{Equations of thermal convection}

Now let us take the well known generalized equations of the thermal convection in Boussinesq approximation to describe the behaviour of the incompressible fluid in a Hele-Shaw cell. This system includes the Navier-Stokes equation, generalized equation of heat transfer and mass conservation law:

$$
\begin{gathered}
\frac{\partial \vec{v}}{\partial t}+(\vec{v} \nabla) \vec{v}=-\frac{1}{\rho} \nabla p+v \Delta \vec{v}+\mathrm{g} \beta T \vec{\gamma}, \\
\frac{\partial T}{\partial t}+(\vec{v} \nabla) T=\operatorname{div}(\chi(T) \nabla T), \quad \operatorname{div} \vec{v}=0 .
\end{gathered}
$$

Here $\vec{v}, p$ and $T$ are dimensional fields of velocity, pressure and temperature. Parameters $v$ and $\chi$ are the coefficients of kinematic viscosity and thermal diffusivity, respectively, $\beta$ is the coefficient of thermal expansion, $\rho$ is the average density of the fluid, $g$ is the gravitational acceleration, and $\vec{\gamma}$ is the unit vector oriented vertically upward. We assume the viscosity to be constant to simplify our problem. On the other hand, thermal diffusivity is determined by

$$
\chi=\kappa /\left(\rho C_{p}\right),
$$

where $\kappa$ and $\mathrm{C}_{\mathrm{p}}$ are the coefficient of thermal conductivity and heat capacity, respectively. The dependence of $\chi$ on temperature like in (2.2) comes from treating the density and thermal capacity as constants.

Let us summarize experimental data $[5,6]$ and suppose that the dependence of thermal diffusivity on temperature can be expressed by the simple formula:

$$
\chi(T)=\chi_{o}(1+\alpha T) .
$$

Here $T$ is the deviation of temperature on conventional convective zero, $\alpha$ is dimensional material coefficient which characterizes the dependence of thermal diffusivity on temperature. Theoretically it can be negative as well as positive. If $\alpha$ is negative, the full temperature additive in formula (2.3) must be sufficiently small and have to be less than the unit. Note that thermal diffusivity is essentially a positive parameter.

The sides of the cavity are supposed to be rigid, therefore the velocity vanishes on the boundaries. As a result of the heating from below the specific temperature distribution takes place on the boundary. Namely, the unknown velocity and temperature fields in equations (2.1), (2.2) must satisfy to the following boundary conditions:

$$
\left.\vec{v}\right|_{\Gamma}=0,\left.\quad T\right|_{\Gamma}=f(x, y) .
$$

These relations account for the existence of no slip condition on all sides of the cell. Furthermore, the cavity has boundaries with high thermal conductivity.

Let us fulfil the numerical analysis of the problem (2.1), (2.2) in the terms of non-dimensional variables. We shall use following set of units during the simulation:

$$
\begin{array}{ll}
\text { length } & {[x, y, z]-d ;} \\
\text { time } & {[t]-d^{2} / v ;} \\
\text { velocity } & {[v]-\chi_{0} / d ;} \\
\text { temperature } & {[T]-\Theta_{i}} \\
\text { pressure } & {[p]-\rho v \chi_{0} / d^{2} .}
\end{array}
$$

We use the ordinary conventions in thermal convection: the viscous and thermal conductive units to measure the time and velocity, respectively. The equations of thermal convection in non-dimensional form can be written as

$$
\begin{gathered}
\frac{\partial \vec{v}}{\partial t}+\frac{1}{\operatorname{Pr}}(\vec{v} \nabla) \vec{v}=-\nabla p+\Delta \vec{v}+\operatorname{Ra} T \vec{\gamma}, \\
\operatorname{Pr} \frac{\partial T}{\partial t}+(\vec{v} \nabla) T=\operatorname{div}(\chi(T) \nabla T), \quad \operatorname{div} \vec{v}=0, \\
\chi(T)=1+\varepsilon T .
\end{gathered}
$$

There are three governing parameters in equations system $(2.4)-(2.6)$

$$
\mathrm{Ra}=\mathrm{g} \beta \Theta d^{3} / v \chi_{0}, \quad \operatorname{Pr}=v / \chi_{0}, \quad \varepsilon=\alpha \Theta .
$$

First and second parameters are the Rayleigh and Prandtl numbers, correspondingly. Non-dimensional parameter $\varepsilon$ describes the dependence of thermal diffusivity on temperature.

In our statement the conditions on the upper and lower boundaries of the Hele-Shaw cell for the nondimensional fields of velocity and temperature have the form:

$$
\begin{array}{ll}
y=0: & \vec{v}=0, T=1 \\
y=H: & \vec{v}=0, T=0
\end{array}
$$

Note that in experiment the variation of Rayleigh number implies usually the change of temperature difference on the upper and lower boundaries. But parameter $\varepsilon$ also depends on $\Theta$. Thus the growth of Rayleigh number for the fixed $\varepsilon$ must be interpreted as the increase of the buoyancy force in comparison with the effect of thermal diffusivity dependence on temperature for the constant heating.

In addition it is important to estimate the value of parameter $\varepsilon$ for any typical case. Thermal diffusivity and kinematic viscosity for metal melts can be evaluated as $\chi \sim 10^{-6} \mathrm{~m}^{2} / \mathrm{s}, v \sim 10^{-7} \mathrm{~m}^{2} / \mathrm{s}\left(\operatorname{Pr} \sim 10^{-2}\right)$ and $\beta \sim$ $10^{-4} 1 / \mathrm{K}$. For the value of Rayleigh number $3 \cdot 10^{2}$ (over-criticality is equal to $1.07$ in Hele-Shaw cell with aspect ratio for wide boundaries 20:40) and thickness $3 \mathrm{~mm}(d=1.5 \mathrm{~mm})$ the character temperature difference on the heat exchangers is estimated as $4.4 \mathrm{~K}$. Thus, for $\alpha \sim 0.051 / \mathrm{K}$ [5] non-dimensional parameter $\varepsilon$ is evaluated as $0.22$.

\subsection{Mechanical equilibrium state}

The mechanical equilibrium state takes place for small values of Rayleigh number. Mechanical equilibrium is characterized by the following conditions:

$$
\partial / \partial t=0, \quad \vec{v}=0 .
$$

As a result, the equations system $(2.4),(2.5)$ turns into

$$
\nabla T_{O} \times \vec{\gamma}=0, \quad \operatorname{div}\left\{\chi\left(T_{O}\right) \nabla T_{O}\right\}=0
$$
\begin{center}
\includegraphics[scale=0.5]{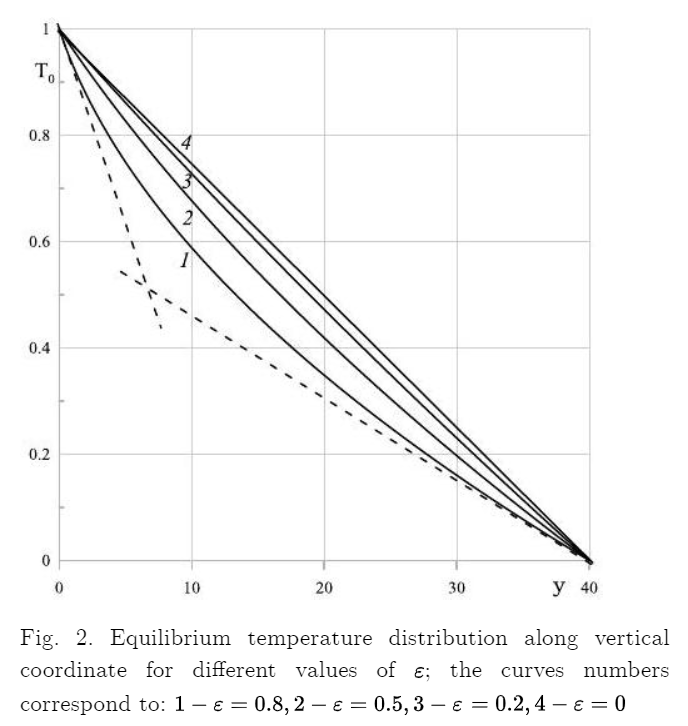}
\end{center}

The exact solution of these equations has to be found taking into account the expression (2.6) for thermal diffusivity and boundary conditions for tem- perature. For different signs of the $\varepsilon$ the solution can be written in following form:

$$
T_{0}=\mp \frac{1}{\varepsilon} \pm \sqrt{\left(1 \pm \frac{1}{\varepsilon}\right)^{2}-\left(1 \pm \frac{2}{\varepsilon}\right) \frac{y}{H}}
$$

where $H$ is the non-dimensional height. The lower mathematical signs correspond to negative values of $\varepsilon_{i}$ upper signs describe the temperature distribution for positive $\varepsilon$.

The curves characterizing the temperature distribution in dependence on vertical coordinate are presented in Fig. 2 for different values of $\varepsilon$.

In spite of linearity of the expression (2.6) for $\chi(T)$ the profile of the temperature is appreciably nonlinear in the state of mechanical equilibrium. Nevertheless, it is easy to see that the dependence becomes linear in the limiting case $\varepsilon=0$.

\section{Linear stability problem}

Let us divide the full fields of velocity, temperature and pressure into the two parts which correspond to basic state (index 0) and small non-stationary perturbations (prime):

$$
\vec{v}=\vec{v}^{\prime}, \quad p=p_{o}+p^{\prime}, \quad T=T_{o}+T^{\prime} .
$$

Equilibrium state in Hele-Shaw cell is considered as a basic one therefore $\vec{v}_{o}=0$. Let us substitute (3.1) into the starting equations (2.4), (2.5) and linearize them. The resulting evolutionary equations for small disturbances have the form:

$$
\partial \vec{v} / \partial t=-\nabla p+\Delta \vec{v}+\operatorname{Ra} T \vec{\gamma},
$$

$$
\operatorname{Pr} \partial T / \partial t+(\vec{v} \nabla) T_{0}=
$$

$$
=\varepsilon \operatorname{div}\left\{T \nabla T_{0}\right\}+\operatorname{div}\left\{\left(1+\varepsilon T_{0}\right) \nabla T\right\} \text {, }
$$

$$
\operatorname{div} \vec{v}=0 \text {. }
$$

Here $\vec{v}, p$ and $T$ are the perturbations of velocity, pressure and temperature, correspondingly. The lateral vertical sides of the cavity are isothermal. In addition on upper and lower heat exchangers the boundary conditions for unknown fields of the velocity and deviation of temperature from equilibrium profile have to be written as

$$
y=0, H: \quad \vec{v}=0, T=0
$$

Preliminarily, it is convenient to rewrite initial differential equations (3.2)-(3.4) in the terms of temperature $T$ and stream function $\psi$. The solution of the boundary-value problem (3.2)-(3.4), (3.5) has the form of normal monotonous modes:

$$
T, \psi \sim(\vartheta, \sigma) e^{-\lambda t} e^{i n \pi x / L} \cos \left(\frac{\pi z}{2}\right)
$$

Here $\vartheta, \sigma$ are the amplitudes of the respective perturbations; $L$ is non-dimensional length, $\lambda$ is an exponential decay rate, $n$ is the integer value that characterizes the periodicity of the perturbation along the $x$ axis. The stream function is connected with the components of velocity by means of relations

$$
v_{x}=\partial \psi / \partial y, v_{y}=-\partial \psi / \partial x .
$$

There are two distinct cases: $\varepsilon$ is positive or negative. Later we will only examine the case with $\varepsilon>0$ in our calculations that means the growth of thermal diffusivity with the temperature.

In the following stage we assume the decrement to be equal zero. The perturbations of this type do not grow and do not decrease. These perturbations are called by the neutral one. In this case Rayleigh number Ra plays the role of the eigenvalue in the spectral amplitude problem. It depends on all parameters like $H$, $L, n$ and $\varepsilon$. It should be emphasized that the linear problem for small disturbances does not contain the Prandtl number. So the boundary of stability does not depend on this parameter.

\subsection{Limiting case of $\varepsilon=0$}

At first let us consider the limiting case of $\varepsilon=0$ and find analytically the expression for neutral curves i.e. the function of critical Rayleigh number in dependence on the parameters of the problem. Setting the exponential decay rate $\lambda$ to zero, we shall be interested in neutral disturbances. If dependence of thermal diffusivity on temperature is absent, the temperature distribution along vertical is linear and the equations system for amplitudes $\xi$ and $\theta$ of small perturbations can be represented in the following form [3]:

$$
\begin{aligned}
&\frac{1}{H} \frac{\partial \xi}{\partial x}=\frac{\partial^{2} \theta}{\partial x^{2}}+\frac{\partial^{2} \theta}{\partial y^{2}}-\frac{\pi^{2}}{4} \theta, \\
&\Delta_{1}^{2} \xi-\frac{\pi^{2}}{4} \Delta_{1} \xi-\operatorname{Ra} \frac{\partial \theta}{\partial x}=0 .
\end{aligned}
$$

Here $\Delta_{1}$ is the plane Laplacian. We shall search the solution of these equations in the form of simple harmonics

$$
\begin{aligned}
&\xi=a \sin (\pi n x / L) \sin (\pi m y / H) \\
&\theta=b \cos (\pi n x / L) \sin (\pi m y / H)
\end{aligned}
$$

where $n=1,2,3 \ldots, m=1,2,3 \ldots$ After the substitution of these functions into the equations system (3.8), (3.9) the dispersion relation gives:

$$
\mathrm{Ra}_{\mathrm{c}}=\frac{\pi^{4} \mathrm{~L}^{2} \mathrm{H}}{\mathrm{n}^{2}}\left(\frac{\mathrm{m}^{2}}{\mathrm{H}^{2}}+\frac{\mathrm{n}^{2}}{\mathrm{~L}^{2}}\right)\left(\frac{\mathrm{m}^{2}}{\mathrm{H}^{2}}+\frac{\mathrm{n}^{2}}{\mathrm{~L}^{2}}+\frac{1}{4}\right)^{2}
$$

The analysis of this formula for $L=20, H=40$ shows that the three lower values of the Rayleigh number in our spectrum correspond to the modes $(2,1),(3,1)$ and $(1,1)$ (here the indexes describe the periodicity of the flow along axes $x, y$ and conform to the different values of $n$ and $m$, respectively).

\subsection{General case of arbitrary $\varepsilon$}

In general case for arbitrary values of parameter $\varepsilon$ the calculations were fulfilled by means of Galerkin method $[8,9]$. The coefficients in equations (3.2), (3.3) depend on only $y$. So the solution on $x$ can be found in the form of simple harmonics (3.6). As a result the system of two ordinary differential equations with variable coefficients for amplitudes $\sigma$ and $\vartheta$ takes place

$$
\begin{gathered}
-\frac{\pi n}{L} \frac{d T_{o}}{d y} \sigma=2 \varepsilon \frac{d T_{o}}{d y} \vartheta^{\prime}+\varepsilon \frac{d^{2} T_{o}}{d y^{2}} \vartheta+ \\
+\left(1+\varepsilon T_{o}\right)\left(\vartheta^{\prime \prime}-\left(\frac{\pi n}{L}\right)^{2} \vartheta-\frac{\pi^{2}}{4} \vartheta\right), \\
\sigma^{I V}-2\left(\frac{\pi n}{L}\right)^{2} \sigma^{\prime \prime}+\left(\frac{\pi n}{L}\right)^{4} \sigma- \\
-\frac{\pi^{2}}{4}\left(\sigma^{\prime \prime}-\left(\frac{\pi n}{L}\right)^{2} \sigma\right)+R a \frac{\pi n}{L} \vartheta=0 .
\end{gathered}
$$

In a limiting case of $\varepsilon=0$ the analysis of (3.8), (3.9) leads to important result that the three most dangerous mode for the cavity with the aspect ratio 2:20:40 are $(2,1),(3,1)$ and $(1,1)$.

Thus, let us apply the Galerkin's procedure for both unknown functions $\sigma$ and $\theta$ with only one basis function: $(\sigma, \vartheta) \sim \sin (\pi y / H)$.

Termwise integration of (3.11), (3.12) with the weight was carried out numerically. The resulting data for critical Rayleigh numbers in dependence on parameter $\varepsilon$ are presented in Fig. 3 in the case of a cavity with $L=20, H=40$. As parameter $\varepsilon$ increases, the critical Rayleigh number increased too: when $\varepsilon=0$, $\mathrm{Ra}_{\mathrm{c}}=281$ and when $\varepsilon=0.8, \mathrm{Ra}_{\mathrm{c}}=417$.
\begin{center}
\includegraphics[scale=0.5]{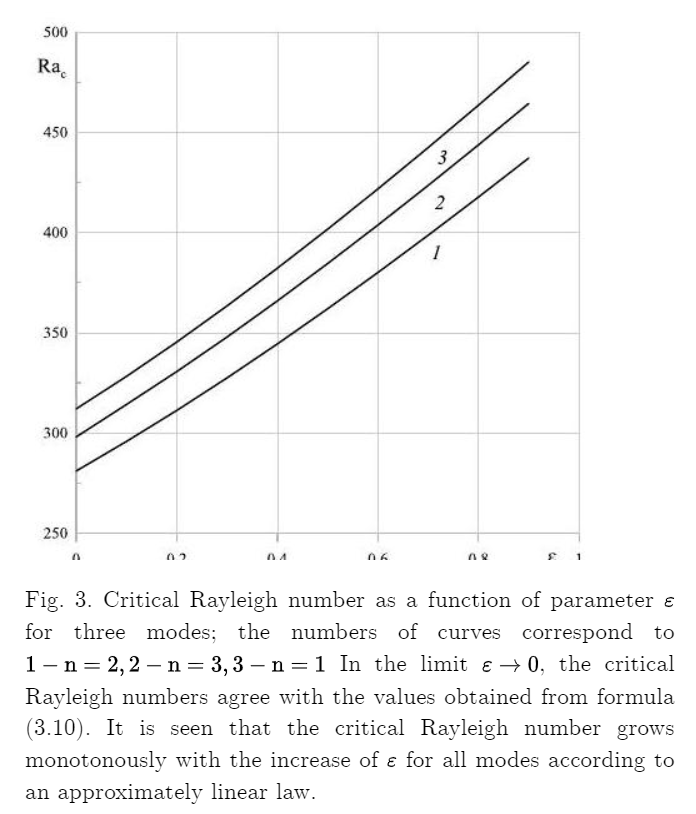}
\end{center}

\section{Non-linear regimes of convection}

The geometry of the problem permits to use the approximation of plane trajectories. As a result it is convenient to execute numerical simulation on the basis of the full non-linear equations written in terms of stream function and vorticity. Let us exclude the pressure and introduce stream function by the same way as in the course of the solution of the linear problem. Finally the standard system of non-linear equations in terms of the stream function $\psi$, vorticity $\varphi$ and temperature $T$ has the form

$$
\begin{gathered}
\frac{\partial \varphi}{\partial t}+\frac{1}{\operatorname{Pr}}\left(\frac{\partial \varphi}{\partial x} \frac{\partial \psi}{\partial y}-\frac{\partial \varphi}{\partial y} \frac{\partial \psi}{\partial x}\right)=\Delta \varphi-\operatorname{Ra} \frac{\partial T}{\partial x}, \\
\frac{\partial T}{\partial t}+(\vec{v} \nabla) T=\operatorname{div}(\chi(T) \nabla T), \varphi=\Delta_{1} \psi .
\end{gathered}
$$

Boundary conditions for equations system (4.1), (4.2) have to be written in the form:

$$
\begin{aligned}
&x=0, \mathrm{~L}: \quad \psi=\psi y^{\prime}=0, \mathrm{~T}=\mathrm{T}_{0}(y) ; \\
&y=0, \mathrm{H}: \quad \psi=\psi_{x}{ }^{\prime}=0, \mathrm{~T}=1,0 .
\end{aligned}
$$

It is convenient to divide the full temperature field on equilibrium part $T_{o}(y)$ and deviation $T_{1}(x, y, t)$. The small thickness of the cavity permits to impose the certain form of the solution depending on the transversal coordinate $z$ and apply the Galerkin procedure. Let us separate the variables and search the solution as

$$
\begin{aligned}
&\psi=\sigma(x, y, t) \cos (\pi z / 2), \\
&T_{1}=\vartheta(x, y, t) \cos (\pi z / 2),
\end{aligned}
$$

where the $\sigma, \vartheta$ are the amplitudes of stream function and temperature which depend on time and coordinates in the plane of wide boundaries. After the averaging on $z$ with the corresponding weight the equations for amplitudes can be written as

$$
\begin{gathered}
\frac{\partial \phi}{\partial t}+\frac{8}{3 \pi \operatorname{Pr}}\left(\frac{\partial \phi}{\partial x} \frac{\partial \sigma}{\partial y}-\frac{\partial \phi}{\partial y} \frac{\partial \sigma}{\partial x}\right)= \\
=\Delta_{1} \phi-\frac{\pi^{2}}{4} \phi-R a \frac{\partial \vartheta}{\partial x} \\
\operatorname{Pr} \frac{\partial \vartheta}{\partial t}+\frac{8}{3 \pi}\left(\frac{\partial \vartheta}{\partial x} \frac{\partial \sigma}{\partial y}-\frac{\partial \vartheta}{\partial y} \frac{\partial \sigma}{\partial x}\right)-\frac{d T_{o}}{d y} \frac{\partial \sigma}{\partial x}= \\
=2 \varepsilon \frac{d T_{o}}{d y} \frac{\partial \vartheta}{\partial y}+\left(1+\varepsilon T_{o}\right)\left(\Delta_{1} \vartheta-\frac{\pi^{2}}{4} \vartheta\right)+ \\
+\varepsilon \frac{d^{2} T_{o}}{d y^{2}} \vartheta+\frac{8 \varepsilon}{3 \pi}\left(\Delta_{1} \vartheta-\frac{\pi^{2}}{4} \vartheta\right) \vartheta+\frac{\varepsilon \pi}{3} \vartheta^{2}+
\end{gathered}
$$

$$
+\frac{8 \varepsilon}{3 \pi}\left(\left(\frac{\partial \vartheta}{\partial x}\right)^{2}+\left(\frac{\partial \vartheta}{\partial y}\right)^{2}\right)
$$

Boundary conditions for equations system (4.3), (4.4) have to be rewritten in the form:

$$
\begin{array}{ll}
x=0, \mathrm{~L}: & \sigma=\sigma_{y^{\prime}}=0, \vartheta=0 ; \\
y=0, \mathrm{H}: & \sigma=\sigma_{x^{\prime}}=0, \vartheta=0 .
\end{array}
$$

\subsection{Description of numerical procedure}
\begin{center}
\includegraphics[scale=0.5]{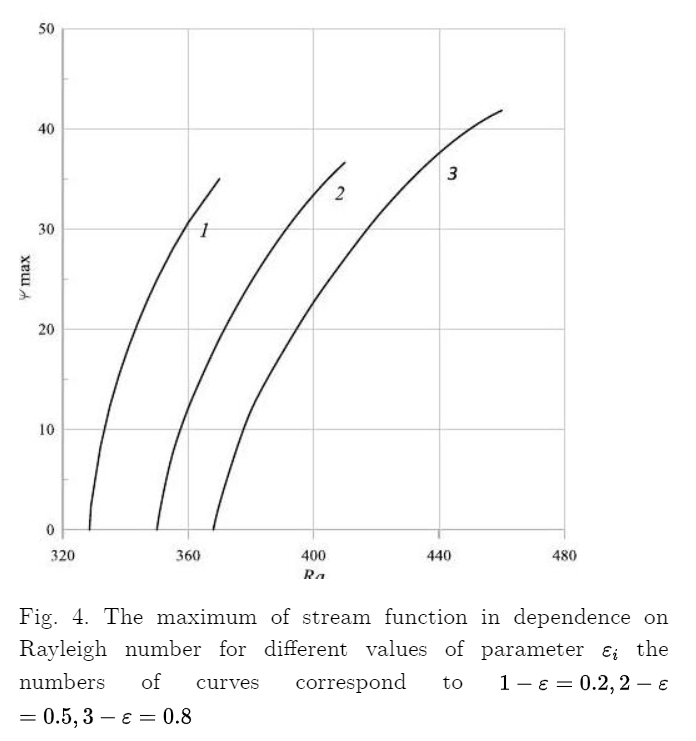}
\end{center}
The method of finite differences was applied to calculate over-critical regimes. Numerical code was written in programming language Fortran- 90 . The problem was considered in terms of the vorticity and stream function i.e. the so-called two-fields method of the solution was used. Explicit scheme was realized to simulate the dynamics of convective system. The basic mesh contained 37:31 nodes. The first order coordinate derivatives were approximated by the central differences with the second order accuracy. The time derivative was expressed by the one-sided difference with the first order accuracy. The Laplace operator was factorized on the base of the three point scheme and had the second order accuracy. The values of vorticity on the lateral boundaries were found over the formula of Thom and Aplte [10]. The step size of the time was calculated in accordance with the necessity of the stability of numerical procedure over the formula

$$
\tau=\min \left(h_{x}^{2}, h_{y}^{2}\right) / 4 \delta,
$$

where $h_{x}, h_{y}$ are the coordinate steps on $x$ and $y, \delta$ is the empirical parameter greater than unit. The Poisson equation for the stream function $\psi$ was solved by the 
\begin{center}
\includegraphics[scale=0.7]{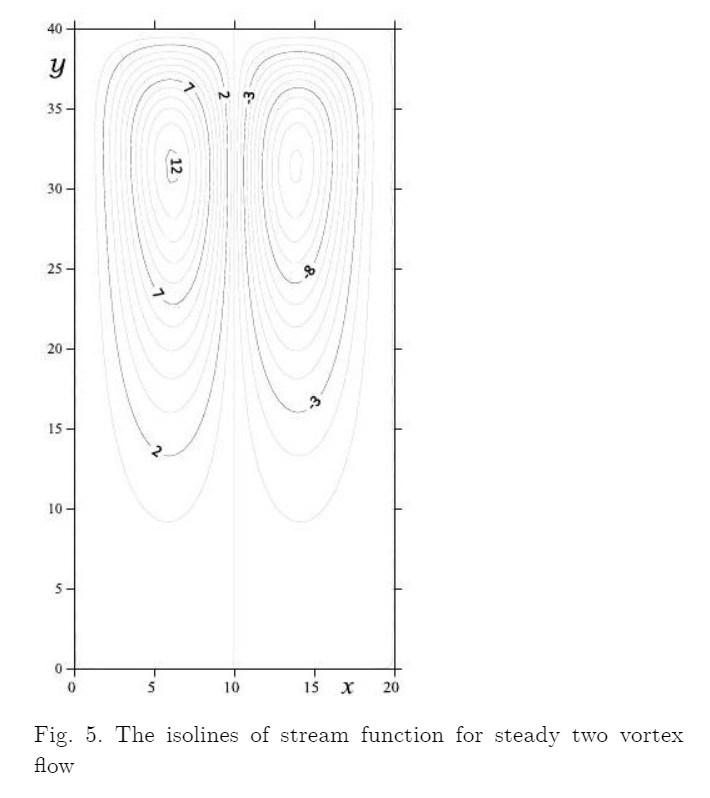}
\end{center}

method of simple iterations [10]. Over the numerical simulation pseudoviscosity method was used to get the snap fields of the temperature $T$ and the stream function $\psi$. To analyze the oscillatory regimes maximum and minimum values of stream function $\psi_{\max }$ and $\psi_{\min }$ were calculated.

\subsection{Results and discussion}

The numerical calculations with isothermal wide boundaries of the Hele-Shaw cell yielded the following results. The first over-critical regime with the lowest value of Ra was found to correspond to the characteristic two vortex flow. This result was independent of the value of $\varepsilon$. For $\varepsilon=0.2$, the relation between the flow regime and Ra was investigated in detail.

At low Rayleigh numbers below $\mathrm{Ra}=330$, a stationary equilibrium state flow was observed. As Ra was increased, we found a two vortex flow regime with left-right symmetry but this flow had no up-down symmetry. The isolines of stream function and temperature are shown in Figs. 5,6 for $\mathrm{Ra}=360, \operatorname{Pr}=7$, and $\varepsilon=0.5$. The displacement of vortices to the upper part of the cavity is observed. There is a simple explanation of this effect.

The bowed equilibrium profile of temperature can be replaced approximately by the broken line with two straight pieces (dashed lines, Fig. 2). There are two regions in a cavity at the plane of wide boundaries with different characteristic values of temperature gradient. For small values of super-criticality the higher part of temperature distribution with large derivative induces convection but the lower part of the broken line may
\begin{center}
\includegraphics[scale=0.7]{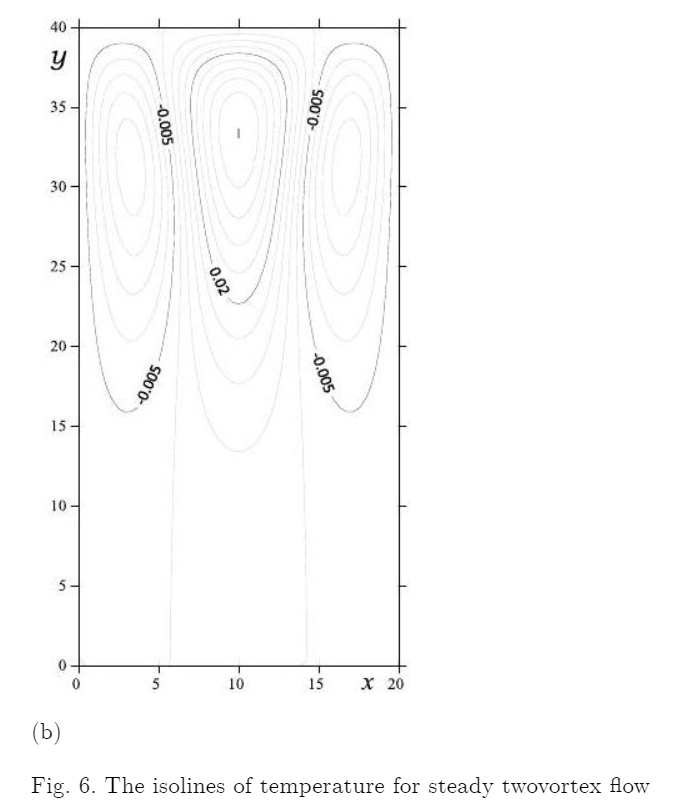}
\end{center}
correspond to the small temperature gradient which is not sufficient to produce convective motion. Therefore the stagnant zone has been formed in lower part of the Hele-Shaw cell. The space with the slow movement can be observed from the fields of stream function and temperature (Figs. 5, 6).

At $\mathrm{Ra}=455$ and beyond, we found an oscillatory four vortex flow with reunification of corner vortices. Initially the oscillations in this regime were periodic and contained only one frequency at $\mathrm{Ra}=455$. However in the range $460<\mathrm{Ra}<500$, the dependence of $\psi_{\max }$ on time clearly contained more frequencies and a Fourier analysis is required to understand these complexities. We also looked at the dependence of $\psi_{\max }$ on Ra with different values of parameter $\varepsilon$ for flows over the threshold of convection. It can be seen that for all values of $\varepsilon, \psi_{\max }$ increases with the Rayleigh number.

The x-intercepts of these graph had to be extrapolated from the graphs as calculating the value of $\psi_{\max }$ close to zero would have large errors induced by our numerical approach for big values of $\varepsilon$. Furthermore, the $\mathrm{x}$-intercepts obtained here are close to the projections that can be obtained from Fig. 3. The difference in the two values obtained is due to the error introduced by the using of only one basis function in $\mathrm{Ga}-$ lerkin's method.

If we introduce point perturbation at the initial moment of time, the two-vortex steady regime is characterised by the down flow along lateral sides and upward movement along centerline. This flow gives the thermal spot in central part of the cavity (Fig. 6). 
\begin{center}
\includegraphics[scale=0.8]{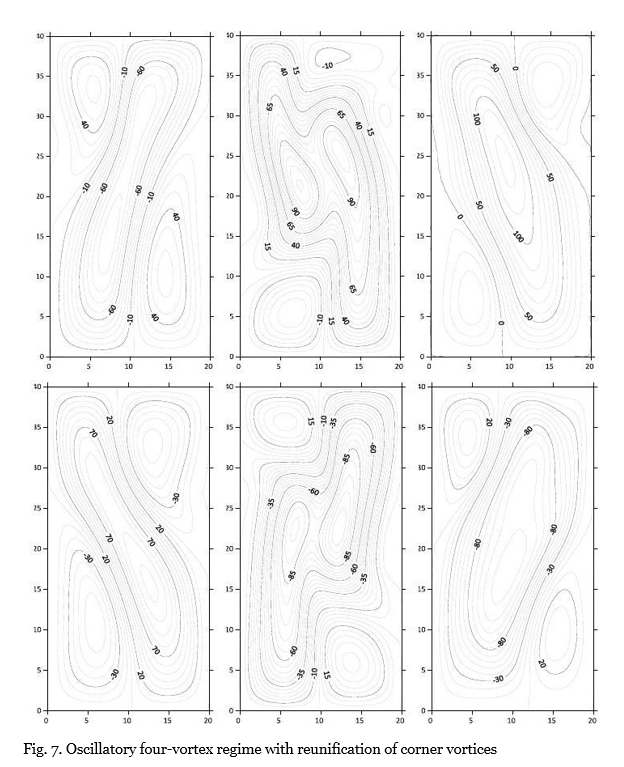}
\end{center}
The period $\tau$ of oscillations for different values of Rayleigh number

\begin{tabular}{|c|c|c|c|c|}
\hline $\mathrm{Ra}$ & 455 & 460 & 465 & 470 \\
\hline$\tau$ (non-dimensional units) & 398 & 392 & 360 & 282 \\
\hline$\tau$ (minutes) & $14.9$ & $14.7$ & $13.5$ & $10.6$ \\
\hline
\end{tabular}

It was emphasized that oscillatory regimes originate for moderate values of super-criticality. In wide range of governing parameter Ra there are different variations of the four vortex flow with alternating reunification of corner vortices. Characteristic fields of stream function in different moments time are presented in Fig. 7 for $\mathrm{Ra}=500$. The Fig. 8 demonstrates that the oscillations become more non-linear and their amplitude increases with the growth of governing parameter Ra. At the same time the period of oscillations has the tendency to be smaller with the increase of Rayleigh number. The numerical values of period can be found in the table. For large values of super-criticality this reunification of corner vortices becomes nonperiodical.\\

\section{Conclusion}

Convective flows in a Hele-Shaw cell have been investigated theoretically when the dependence of thermal diffusivity on temperature is taken into account. It has been shown for the cavity with aspect ratio $2: 20: 40$ that the inclusion of this factor in our model leads to the symmetry breakdown of the steady twovortex flow for small values of over-criticality. Visually it looks as the displacement of vortices to upper part of the cavity and formation of the stagnation zone near the lower boundary. The different oscillatory flows originate for bigger values of Rayleigh number. It was found that these oscillatory regimes correspond to various deviations on the well-known four-vortex flow with alternating reunification of opposite corner vortices.\\ 
\begin{center}
\includegraphics[scale=0.8]{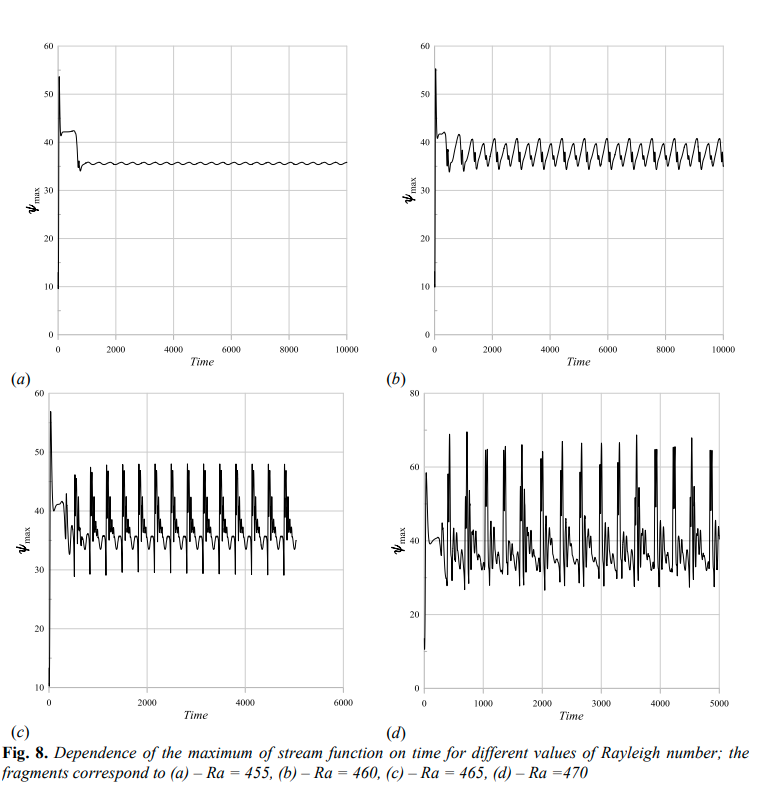}
\end{center}
We are grateful for the financial support of the scholarship provided by the British Petroleum and the University of Oxford's Career Service Office which enabled us to obtain these scientific results.

\section{References}

1. Gershuni G. Z., Zhukhovitskii E. M. Convective stability of incompressible fluids. Jerusalem: Keter Publishing House, 1976. $330 \mathrm{p}$.\\

2. Babushkin I. A., Glazkin I. V., Demin V. A., Platonova A. N., Putin G. F. Variability of a typical flow in a Hele-Shaw cell. Fluid Dynamics, 2009, vol. 44, no. 5, pp. 631-640.\\

3. Demin V. A., Petukhov M. I. The effect of temperature dependence of the viscosity on stationary convective flows in Hele-Shaw cell.\\

5. Savchenko I. V. Experimental'noye issledovaniye teploprovodnosti i temperaturoprovodnosti rasplavov legkoplavkih metallov i splavov metodom lazernoy vspyshki. Avtoreferat kandidatskoy dissertatsii, Institut teplofiziki im. S.S. Kutateladze SO RAN, Novosibirsk, 2011. 20 p (In Russian).\\

6. Solomin B. A., Hodakov A. M. Definition of thermal diffusivity of a multicomponent liquid at it cooling. Izvestia of Samara Scientific Center of the Russian A cademy of Sciences, 2008, vol. 10, no. 3, pp. 716-718.\\

7. Baldina N. O., Demin V. A. Thermal convection in a horizontal fluid layer in the case of thermal conductivity dependence on temperature. Bulletin of Perm University. Physics, 2015, no. 3 (31), pp. 5 12 .\\

8. Fletcher C. A. J. Computational techniques for fluid dynamics. Vol. 1. Springer-Verlag. 1988. 409 p.\\

9. Fletcher C. A. J. Computational techniques for fluid dynamics. Vol. 2. Springer-Verlag. 1988. 484 p.\\

10. Roache P. Computational fluid dynamics. Albuquerque, New Mexico, Hermosa Pub., 1976 . $446 \mathrm{p}$.\\

\section{Please cite this article in English as:}

Nuckchady R. C., Demin V. A. Thermal convection in a Hele-Shaw cell with the dependence of thermal diffusivity on temperature. Bulletin of Perm University. Physics, 2018, no. $3(41), \mathrm{pp} .55-64 .$ doi: 10.17072/19943598-2018-3-55-64
\end{document}